\documentclass[twocolumn,showpacs, showkeys, preprintnumbers,amsmath,amssymb]{revtex4}

\usepackage{graphicx}
\usepackage{dcolumn}
\usepackage{bm}
\usepackage{enumerate}

\begin{document}


\title{Calculating the trap density of states in organic field-effect transistors from experiment: A comparison of different methods}

\author{Wolfgang L. Kalb}%
 \email{kalb@phys.ethz.ch}
\author{Bertram Batlogg}
\affiliation{%
Laboratory for Solid State Physics, ETH Zurich, 8093 Zurich,
Switzerland
}%

\date{\today}

\begin{abstract}
The spectral density of localized states in the band gap of
pentacene (trap DOS) was determined with a pentacene-based thin-film
transistor from measurements of the temperature dependence and
gate-voltage dependence of the contact-corrected field-effect
conductivity. Several analytical methods to calculate the trap DOS
from the measured data were used to clarify, if the different
methods lead to comparable results. We also used computer
simulations to further test the results from the analytical methods.
Most methods predict a trap DOS close to the valence band edge that
can be very well approximated by a single exponential function with
a slope in the range of $50-60$\,meV and a trap density at the
valence band edge of $\approx2\times10^{21}$\,eV$^{-1}$cm$^{-3}$.
Interestingly, the trap DOS is always slightly steeper than
exponential. An important finding is that the choice of the method
to calculate the trap DOS from the measured data can have a
considerable effect on the final result. We identify two specific
simplifying assumptions that lead to significant errors in the trap
DOS. The temperature-dependence of the band mobility should
generally not be neglected. Moreover, the assumption of a constant
effective accumulation layer thickness leads to a significant
underestimation of the slope of the trap DOS.
\end{abstract}

\maketitle

\section{Introduction} \label{section-introduction}

World-wide research on organic field-effect transistors is at a high
level as this new technology is poised to enter the
market.\cite{BragaD2009, SirringhausH2009} An appealing feature of
this technology is that organic semiconductors can be deposited by
thermal evaporation or from solution at low cost on large areas
while keeping the substrates close to room temperature.
Consequently, organic semiconductors are promising candidates for
future flexible and low-cost electronics.

The mobility of charge carriers in organic field-effect transistors
is comparable to the mobility in hydrogenated amorphous silicon
thin-film transistors (1\,cm$^{2}$/Vs) and thus is already adequate
for many applications.\cite{KelleyTW2003, ZhangXH2007, TakeyaJ2007,
McCullochI2006} In addition to a high mobility, useful organic
transistors must have a near zero threshold voltage, a steep
subthreshold swing and a high electrical and environmental
stability. The transistor parameters and stability of organic
field-effect transistors are intimately related to the efficiency of
the charge transport mechanism and the extend of charge carrier
trapping in extrinsic traps. The main scientific challenge thus is
to clarify the nature of the charge transport mechanism and the
microscopic origin of charge carrier traps in organic field-effect
transistors.

Field-effect transistors can be used to determine the underlying
spectral density of localized states in the band gap, i.e. the trap
densities as a function of energy (trap DOS). This has been
extensively done with thin-film transistors (TFT's) employing
amorphous semiconductors or with TFT's based on polycrystalline
silicon.\cite{SpearWE1972, GrunewaldM1980, SpearWE1980,
GrunewaldM1981, WeisfieldRL1981, PowellMJ1981, WeberK1982,
DjamdjiF1987, SchumacherR1988, FortunatoG1986, MiglioratoP1987,
FortunatoG1988, FogliettiV1999} This approach is expected to be of
great value for the understanding of any novel semiconductor in a
field-effect transistor including organic small molecule
semiconductors, polymeric semiconductors or ZnO.\cite{SunJ2008}

The research efforts to calculate the trap DOS from measurements of
organic field-effect transistors have increased only recently. On
the one hand, the density of states function can be calculated from
the linear regime transfer characteristics in a straightforward
fashion with an analytical method.\cite{HorowitzG1995, SchauerF1999,
ZhivkovI1999, HorowitzG2000, LangDV2004, DeAngelisF2006,
DeAngelisF22006, KalbWL2007, KawasakiN2007, SoWY2007, GuoD2007,
KalbWL2008, SoWY2008, LeufgenM2008, VanoniC2009} This approach has
the advantage of giving an unambiguous result but errors may result
from the various simplifying assumptions. On the other hand, a
density of states function can be postulated \textit{a priori} and
the corresponding transistor characteristics can be calculated by
means of a suitable computer program. The density of states function
is then iteratively refined until good agreement between the
measured characteristic and the computer-simulated curve is
achieved.\cite{VoelkelAR2002, ScheinertS2002, SalleoA2004,
KnippD2005, OberhoffD2007, ScheinertS2007, PernstichKP2008}

Several analytical methods have been used to calculate the trap DOS
from transfer characteristics of organic field-effect transistors.
In order to eventually clarify the microscopic origin of charge
carrier traps in organic field-effect transistors, it is highly
desirable to quantitatively compare the results from various
experiments and from various research groups. An important
prerequisite to such a systematic comparison is to test if the
different analytical methods lead to comparable results. We applied
the most frequently used analytical methods to the same set of
measured transfer characteristics. Moreover, we applied a computer
simulation program to determine a trap DOS that leads to simulated
transfer characteristics closely matching the measured data.

In the following we are dealing with pentacene and p-type
conduction. For convenience the charge carriers are called holes
although they may be of pronounced polaronic
character.\cite{SilinshEA1994} Moreover, we use terms such as
valence band edge, band mobility or effective density of extended
states. However in order to apply the calculation methods described
below, we do not necessarily need to have band transport or the
existence of extended states. The calculation methods may be applied
as long as the charge transport can be described by a transport
level with a distribution of localized states below this transport
level (trap-controlled transport).\cite{ArkhipovVI2001}

We begin by summarizing the widely-used analytical description of an
organic field-effect transistor. This description is only valid for
samples with a low trap density and negligible contact resistances.
The trapping and release times are assumed to be much shorter than
the time necessary to measure a transistor characteristic, i.e. we
have no current hysteresis. In Sec.~\ref{section-extraction} we
present the trap DOS calculated with the different methods for a
pentacene TFT with a SiO$_{2}$ gate dielectric, i.e. a sample with a
significantly high trap density. The essential equations of the
different analytical methods are also given in this section along
with specific details about the use of the calculation methods.

\section{\label{section-analyticalFET} Analytical description of an ideal field-effect transistor}

In the linear regime ($|V_{d}|\ll|V_{g}-V_{t}|$), the drain current
$I_{d}$ of an ideal field-effect transistor is given by
\begin{equation} \label{equation-linear}
I_{d} = \frac{W}{L}\mu_{0}C_{i}(V_{g}-V_{t})V_{d},
\end{equation}
where $C_{i}$ is the capacitance per unit area, $W$ and $L$ are the
channel width and length, $V_{g}$ and $V_{d}$ are the gate voltage
and the drain voltage and $\mu_{0}$ is the band mobility which is
independent of gate voltage. Eq.~\ref{equation-linear} predicts a
linear dependence of the drain current on the effective gate voltage
$I_{d}\propto(V_{g}-V_{t})$. A linear regression of the measured
transfer characteristic thus yields the band mobility $\mu_{0}$ and
the threshold voltage $V_{t}$. The threshold voltage is defined as
the gate voltage above which essentially all of the incrementally
added gate-induced charge is mobile (``free''). The threshold
voltage depends on the trap density in the device and on the
flatband voltage $V_{FB}$. The flatband voltage is the gate voltage
which needs to be applied in order to enforce flat bands at the
insulator-semiconductor interface. A non-zero flatband voltage can
result from a difference of the Fermi level in the semiconductor and
in the gate electrode. More importantly, the flatband voltage is
influenced by charge that is permanently trapped at the
insulator-semiconductor interface or within the gate dielectric.

\begin{figure}
\includegraphics[width=0.90\linewidth]{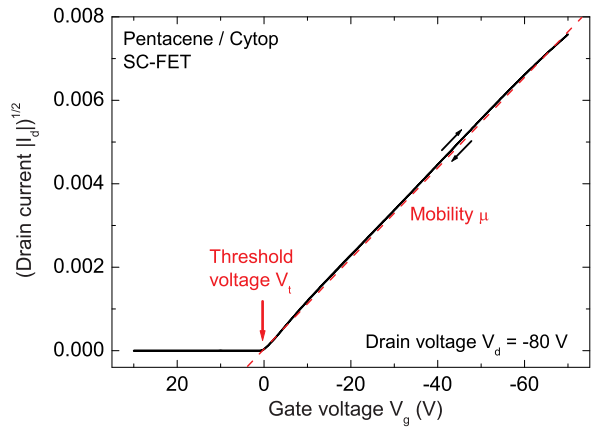}
\caption{\label{figure-ideal2} (Color online) High-performance
pentacene single-crystal field-effect transistor (SC-FET) with a
Cytop$^{TM}$ fluoropolymer gate dielectric. The graph shows the
square root of the drain current (full black line) and the dashed
red line is a linear fit of the measured data. The square root of
the drain current linearly depends on gate voltage in accordance
with the well-known field-effect transistor equation for the
saturation regime. This linear dependence is a mark of the low trap
density at the insulator-semiconductor interface, as well as the
high field-effect mobility of $\mu=1.4$\,cm$^{2}$/Vs and the
near-zero threshold voltage.}
\end{figure}

The drain current in the saturation regime ($|V_{d}| \geq
|V_{g}-V_{t}|$) quadratically depends on gate voltage, i.e.
\begin{equation} \label{equation-saturation}
I_{d}=\frac{W}{L}\frac{\mu_{0} C_{i}}{2}(V_{g}-V_{t})^{2}.
\end{equation}
Fitting a straight line to the square root of the measured drain
current yields the band mobility $\mu_{0}$ and the threshold voltage
$V_{t}$. This ideal behaviour can be observed in organic
field-effect transistors with a low trap density. For example in
Fig.~\ref{figure-ideal2} we show the near-ideal transfer
characteristic of a pentacene single crystal field-effect transistor
(SC-FET) with a Cytop$^{TM}$ fluoropolymer gate
dielectric.\cite{KalbWL2007}

\begin{figure}
\includegraphics[width=0.90\linewidth]{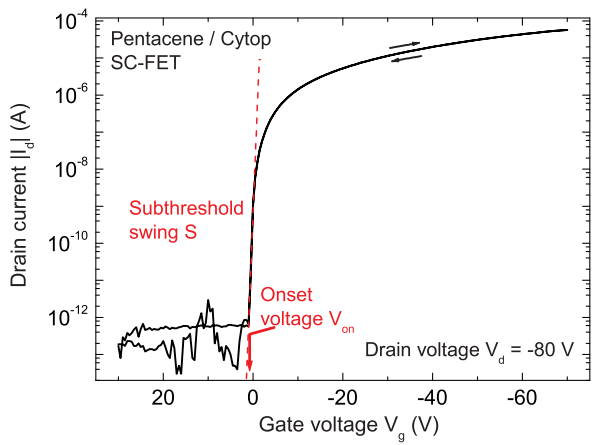}
\caption{\label{figure-ideal1} (Color online) Same data as in
Fig.~\ref{figure-ideal2} of a pentacene/Cytop SC-FET plotted on a
logarithmic scale. The forward and reverse sweeps are shown. The
very low trap density in the active region of the transistor
manifests itself in several desirable properties: near-zero onset
voltage V$_{on}$, very steep subthreshold swing of $S=0.3$\,V/dec,
as estimated from the dashed red line, and negligible current
hysteresis.}
\end{figure}

The onset voltage $V_{on}$ and the subthreshold swing $S$ are other
important device parameters. The onset voltage is defined as the
gate voltage where the drain current exceeds the noise level which
typically is at 10$^{-12}$\,A (see Fig.~\ref{figure-ideal1}). The
subthreshold swing is a measure of how easily a transistor can be
switched from the off-state to the on-state. It is defined as
\cite{SzeSM1981}
\begin{equation} \label{equation-subbasic}
S=\frac{dV_{g}}{d(\log I_{d})}.
\end{equation}
With the simplistic assumption that both the density of deep bulk
traps $N_{bulk}$ and the density of (deep) interface traps $N_{int}$
are independent of energy, the subthreshold swing may be written as
\cite{RollandA1993}
\begin{equation} \label{equation-subground}
S=\frac{kT\ln10}{e}\left[1+\frac{e}{C_{i}}(\sqrt{\epsilon_{s}N_{bulk}}+eN_{int})\right].
\end{equation}
This may be simplified as follows:
\begin{equation} \label{equation-subthreshold}
S=\frac{kT\ln10}{e}\left[1+\frac{e^{2}}{C_{i}}N_{\Box}\right].
\end{equation}
Both the deep bulk traps and the interface traps contribute to the
trap density $N_{\Box}$ (per unit area and unit
energy).\cite{McDowellM2006, YoonMH2006} The subthreshold swing thus
is a simple measure of the deep trap density.
Fig.~\ref{figure-ideal1} shows the same data as in
Fig.~\ref{figure-ideal2} on a logarithmic scale. The subthreshold
swing is as steep as $S=0.3$\,V/dec. With
Eq.~\ref{equation-subthreshold} and $C_{i}=4.3$\,nF/cm$^{2}$ we
calculate a trap density from $S$ as low as
$N_{\Box}=1.1\times10^{11}$\,cm$^{-2}$eV$^{-1}$. Assuming an
effective accumulation layer thickness of $a=7.5$\,nm this results
in a volume trap density of
$N=N_{\Box}/a=1.5\times10^{17}$\,cm$^{-3}$eV$^{-1}$.

\begin{figure}
\includegraphics[width=0.90\linewidth]{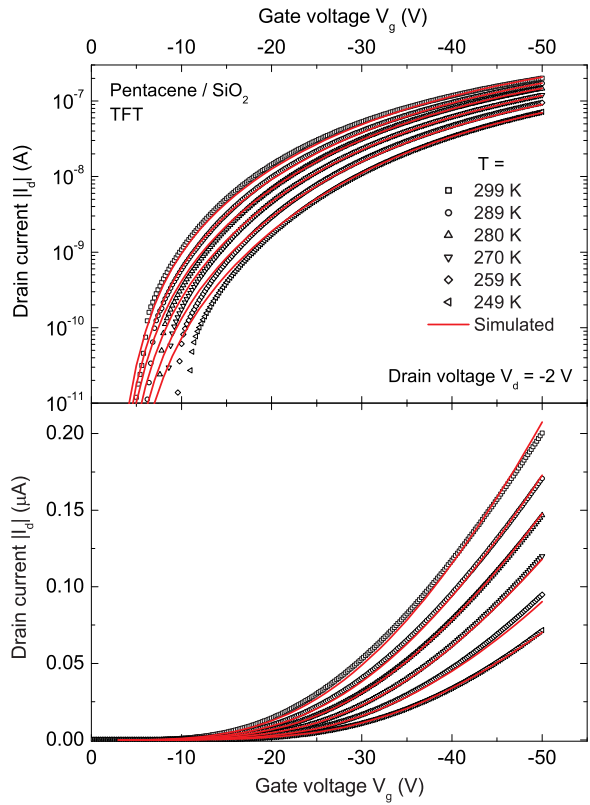}
\caption{\label{figure-tempTFT} (Color online) Transfer
characteristics of a pentacene-based thin-film transistor (TFT) with
a SiO$_{2}$ gate dielectric measured at various temperatures
(symbols). The upper and lower panel show the same data on a
logarithmic and linear scale. The red lines are computer-simulated
transfer characteristics. The increased subthreshold swing (upper
panel) and the reduced field-effect mobility
($\mu\approx0.2$\,cm$^{2}$/Vs at $V_{g}=-50$\,V and $T=299$\,K) are
a result of a relatively high trap density. The high trap density
also results in the drain current to increase faster than linearly
in the linear regime (lower panel).}
\end{figure}

If the experimental transfer characteristics are linear in the
linear regime and quadratic in the saturation regime, the approach
described above is self-consistent and the extracted mobility
$\mu_{0}$ and threshold voltage $V_{t}$ have a clear meaning, i.e.
$\mu_{0}$ is the band mobility and $V_{t}$ is voltage above which
the incrementally added gate-induced charge is placed in the valence
band. In samples with an increased trap density, the drain current
in the linear regime may however increase faster than linearly. For
example, this can be seen in the lower panel of
Fig.~\ref{figure-tempTFT} where we show the measured transfer
characteristics of a pentacene TFT with SiO$_{2}$ gate dielectric
for several temperatures. The transconductance $(\partial
I_{d}/\partial V_{g})_{V_{d}}$ now increases monotonically with gate
voltage. The percentage of the gate-induced holes that are free
increases with gate voltage and this leads to the ``superlinear''
transfer characteristics. Strictly speaking, the threshold voltage
is not reached even at relatively large gate voltages.
Eq.~\ref{equation-linear} and Eq.~\ref{equation-saturation} are not
suitable for this type of transistor.\cite{HorowitzG1999,
HorowitzG2004} This is similar to the case of amorphous silicon
field-effect transistors, where the trap densities are
substantial.\cite{ShurM1984} We note that for the TFT, the gate
capacitance is $C_{i}=13.3$\,nF/cm$^{2}$. This is about three times
larger than the capacitance for the SC-FET in
Fig.~\ref{figure-ideal2} and Fig.~\ref{figure-ideal1}. The
difference between the two types of transistors is even more drastic
than it appears when comparing the graphs.

\section{\label{section-extraction} Quantification of the trap DOS}

Field-effect transistors can be used to quantify the underlying trap
DOS if a more sophisticated description of the device physics is
used. We applied several analytical methods to the linear-regime
transfer characteristics in Fig.~\ref{figure-tempTFT}. We also
applied a simulation program to determine a trap DOS that leads to
simulated transfer characteristics closely matching these measured
transfer characteristics. Moreover, the results were compared to the
crude estimate of the trap density from the subthreshold swing
(Eq.~\ref{equation-subthreshold}). We used a dielectric constant of
$\epsilon_{i}=3.9$ for SiO$_{2}$ and $\epsilon_{s}=3.0$ for
pentacene.\cite{MattersM19992, SchubertM2004} The thickness of the
SiO$_{2}$ gate dielectric was $l=260$\,nm and the pentacene film was
$d=50$\,nm thick. The channel was $L=450$\,$\mu$m long and
$W=1000$\,$\mu$m wide.

All analytical methods and the simulation program are based on the
following simplifying assumptions:
\begin{enumerate}[-]
\item the organic semiconductor is homogenous perpendicular to the
insulator-semiconductor interface, and
\item insulator surface states only introduce an initial band
bending without applied field, i.e. contribute to a non-zero
flatband voltage $V_{FB}$.
\end{enumerate}
As a consequence we obtain an effective trap DOS. In the case of
TFT's with polycrystalline films, the trap densities to be
determined are an average over intra-grain and inter-grain regions
and may also be influenced, to some extend, by trap states on the
surface of the gate dielectric.

\subsection{\label{subsection-analytical}
 Analytical methods}

Several additional assumptions are made to simplify the analytical
methods:
\begin{enumerate}[-]
\item the charge density is homogenous along the transistor
channel (from source to drain), and
\item for the trapped holes, the Fermi function is approximated by
a step function (zero temperature approximation), and
\item the valence band is approximated as a discrete energy level at
the valence band edge $E_{V}$ with an effective density of extended
states $N_{V}$. The occupation of these extended states is
calculated with the Boltzmann function, and
\item the temperature dependence of the Fermi energy $E_{F}$ as well as of the interface potential $V_{0}$ is
neglected (neglect of the statistical shift).
\end{enumerate}
The first assumption is appropriate only if the transfer
characteristics are measured at a low drain voltage. In that case we
can assume the ``unperturbed'' situation where charge is accumulated
by a gate voltage in a metal-insulator-semiconductor (MIS) structure
but no drain voltage is applied.\cite{HorowitzG2004}

\begin{figure*}
\includegraphics[width=0.70\linewidth]{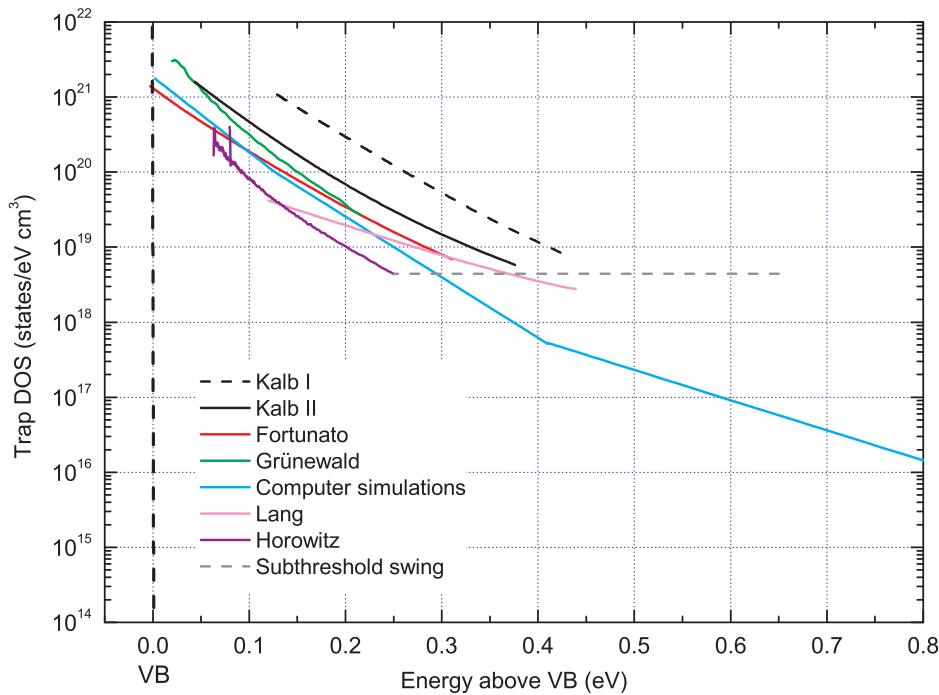}
\caption{\label{figure-compmethods} (Color online) Spectral density
of localized states in the band gap of pentacene (trap DOS) as
calculated with several methods from the same set of measured data.
The energy is relative to the valence band edge (VB). The estimate
from the subthreshold swing (dashed grey line) assumes the trap DOS
not to depend on energy and can only be regarded as a rough estimate
for the traps slightly above the Fermi energy. All other methods
result in a trap DOS that increases slightly faster than
exponentially with energy. The choice of the method to calculate the
trap DOS has a considerable effect on the final result.}
\end{figure*}

The final results from the different methods are summarized in
Fig.~\ref{figure-compmethods} and Table~\ref{table-compmethods}
along with the result from the simulation program. The parameters
$N_{0}$ and $E_{0}$ in Table~\ref{table-compmethods} were obtained
by fitting to each trap DOS an exponential function
\begin{equation}
N(E)=N_{0}\exp(-E/E_{0}).
\end{equation}
Fig.~\ref{figure-compmethods} and Table~\ref{table-compmethods} also
contain the trap density as estimated from the subthreshold swing of
$S=2.4$\,V/dec with Eq.~\ref{equation-subthreshold} and an effective
accumulation layer thickness of $a=7.5$\,nm. Some methods also lead
to an estimate of the band mobility $\mu_{0}$ (see
Table~\ref{table-compmethods}).
\begin{table}
\caption{\label{table-compmethods} Parameters resulting from
different methods to calculate the trap DOS and the trap DOS from
each method is shown in Fig.~\ref{figure-compmethods}. The different
methods were applied to the same set of measured data
(Fig.~\ref{figure-tempTFT}). To obtain the parameters $N_{0}$ and
$E_{0}$, an exponential function $N(E)=N_{0}\exp(-E/E_{0})$ was
fitted to the calculated trap DOS in each case. Most methods lead to
a slope in the range of $E_{0}=50-60$\,meV and to a trap density at
the valence band edge of
$N_{0}\approx2\times10^{21}$\,eV$^{-1}$cm$^{-3}$. Some methods also
lead to an estimate of the band mobility $\mu_{0}$.}
\begin{ruledtabular}
\begin{tabular}{lllll}
Method & Ref. & $N_{0}$ & $E_{0}$ & $\mu_{0}$\footnotemark[1] \\
& & (eV$^{-1}$cm$^{-3}$) & (meV) & (cm$^{2}$/Vs) \\

\hline Kalb I& \cite{KalbWL2008} & $8.5\times10^{21}$ & 60 & 0.7\\
Kalb II & This paper & $2.4\times10^{21}$ & 59 & -\\
Fortunato & \cite{FortunatoG1988, DeAngelisF2006} &
$1.1\times10^{21}$ & 60 & - \\
Gr\"unewald & \cite{GrunewaldM1980, GrunewaldM1981} &
$4.0\times10^{21}$ & 41 & - \\
Computer sim. & \cite{OberhoffD2007} & $1.5\times10^{21}$ & 50 & 0.3\\
Lang\footnotemark[2] & \cite{LangDV2004} & $1.1\times10^{20}$ & 115 & 17 \\
Horowitz & \cite{HorowitzG1995} & $7.0\times10^{20}$ & 48 & 0.2\\
\hline Subthreshold  & \cite{RollandA1993, YoonMH2006, McDowellM2006} & \multicolumn{2}{c}{$N=4.4\times10^{18}$}  & -\\
swing\footnotemark[2]\footnotemark[3] & & & &\\
\end{tabular}
\end{ruledtabular}
\footnotetext[1]{Band mobility at $T=299$\,K.}
\footnotetext[2]{Effective accumulation layer thickness of
$a=7.5$\,nm.} \footnotetext[3]{Assumption: trap DOS independent of
energy.}
\end{table}

The final results in Fig.~\ref{figure-compmethods} and
Table~\ref{table-compmethods} are discussed in
Sec.~\ref{section-discussion}. The simulations are described in more
detail in Sec.~\ref{subsection-computer}. In the following we
describe the different analytical methods and give some specific
details about how the methods were used.

All methods require linear regime transfer characteristics measured
at several temperatures (as in Fig.~\ref{figure-tempTFT}). The
exception is the method by \textit{Gr\"unewald et al.} which only
requires a single linear regime transfer characteristic measured at
one temperature (e.g. room temperature). In order to calculate the
trap DOS from the transfer characteristics $I_{d}(V_{g})$ we need
the field-effect conductivity $\sigma$ and the field-effect mobility
$\mu$ at first. Provided that contact effects are negligible, the
drain current in the linear regime may be written as
\begin{equation} \label{equation-revision1}
I_{d}=\frac{W}{L}\sigma V_{d}
\end{equation}
and the field-effect conductivity $\sigma$ is
\begin{equation}
\sigma=\mu C_{i}(V_{g}-V_{FB}).
\end{equation}
$V_{FB}$ is the flatband voltage and $\mu$ is the (gate-voltage
dependent) field-effect mobility, i.e. an effective mobility in
contrast to the band mobility $\mu_{0}$. The field-effect
conductivity can be calculated from
\begin{equation} \label{equation-sig1}
\sigma(V_{g})=\frac{L}{W}\frac{I_{d}}{V_{d}}.
\end{equation}
The calculation of the field-effect mobility from measurements of
organic field-effect transistors is still controversial for samples
with an increased trap density.\cite{HorowitzG2004} If we
differentiate Eq.~\ref{equation-revision1} with respect to gate
voltage, we have \cite{MottaghiM2006}
\begin{equation} \label{equation-revision2}
\frac{\partial I_{d}}{\partial
V_{g}}=\frac{W}{L}C_{i}V_{d}\left(\mu+(V_{g}-V_{FB})\frac{\partial\mu}{\partial
V_{g}}\right).
\end{equation}
The field-effect mobility is most often calculated from
\begin{equation} \label{equation-mu1}
\mu(V_{g})=\frac{L}{WV_{d}C_{i}}\bigg(\frac{\partial I_{d}}{\partial
V_{g}}\bigg)_{V_{d}}
\end{equation}
which means that the second term in Eq.~\ref{equation-revision2} is
generally neglected.

Contact effects can introduce significant errors when the
field-effect conductivity and the field-effect mobility are
calculated.\cite{TakeyaJ2003, PesaventoPV2004} The contact-corrected
field-effect conductivity can be calculated from gated four-terminal
measurements according to
\begin{equation} \label{equation-sig2}
\sigma(V_{g})=\frac{L'}{W}\frac{I_{d}}{V_{d}'}.
\end{equation}
$L'$ is the distance between the voltage sensing electrodes and
$V_{d}'=V_{1}-V_{2}$ is the voltage drop between these
electrodes.\cite{KalbWL2007, KalbWL2008} The effective field-effect
mobility $\mu$ is not influenced by parasitic contact resistances
when calculated from gated four-terminal measurements according to
\begin{equation} \label{equation-mu2}
\mu(V_{g})=\frac{L'}{WV_{d}'C_{i}}\bigg(\frac{\partial
I_{d}}{\partial V_{g}}\bigg)_{V_{d}}.
\end{equation}

The mobilities as calculated with Eq.~\ref{equation-mu1} or
Eq.~\ref{equation-mu2} overestimate the true field-effect mobilities
to some extend.\cite{MottaghiM2006} This is because for
trap-controlled transport, the mobility increases with gate voltage.
This leads to a positive second term in Eq.~\ref{equation-revision2}
which is neglected in Eq.~\ref{equation-mu1} (and
Eq.~\ref{equation-mu2}). However, the use of Eq.~\ref{equation-mu1}
(or Eq.~\ref{equation-mu2}) is advantageous because the definition
and extraction of a flatband voltage is circumvented. Moreover, this
approach to the field-effect mobility is most often used which
guarantees a good comparability of the mobility values.
Consequently, we choose Eq.~\ref{equation-mu2} (the
contact-corrected version of Eq.~\ref{equation-mu1}) to calculate
the field-effect mobility in the present study. Alternatively, the
field-effect mobility could also be calculated from
\cite{MottaghiM2006}
\begin{equation} \label{equation-revision3}
\mu=\frac{L'}{WV_{d}'C_{i}}\frac{I_{d}}{(V_{g}-V_{FB})}
\end{equation}
In that case we would not need to differentiate the measured data
but a reliable estimate of the flatband voltage (threshold voltage)
would be required.\cite{MottaghiM2006}

In Fig.~\ref{figure-tempTFT} we show the measured transfer
characteristics $I_{d}(V_{g})$ from gated four-terminal
measurements. In addition to the drain current $I_{d}(V_{g})$, the
potentials $V_{1}(V_{g})$ and $V_{2}(V_{g})$ between the grounded
source electrode and the respective voltage sensing electrode were
measured simultaneously while keeping the source-drain voltage
constant. This was done by connecting the source of the transistor
to the ground connector of an HP 4155A parameter analyzer and by
measuring the channel potentials $V_{1}$ and $V_{2}$ with two
additional SMU's in the ``source current - measure voltage'' mode
with a sourced current of 0\,A. For all analytical methods we used
the four-terminal conductivity $\sigma=\sigma(V_{g})$ as derived
from gated four-terminal measurements with Eq.~\ref{equation-sig2}
and the field-effect mobility was calculated according to
Eq.~\ref{equation-mu2}. This allowed for the calculation of a trap
DOS that is free from contact artifacts.\cite{KalbWL2007,
KalbWL2008} Moreover, we only used currents above $|I_{d}|=1$\,nA
for the calculation of trap DOS (1\,nA limit).\cite{KalbWL2008}

For the following description of the analytical methods, Fig.~2 and
Fig.~3 in Ref.~\onlinecite{KalbWL2008} are useful.

\subsubsection{Method by Lang et al.}

For this method (Ref.~\onlinecite{LangDV2004}), the activation
energy $E_{a}(V_{g})$ is defined by
\begin{equation} \label{equation-la0}
\sigma(V_{g})=A\exp\left(-\frac{E_{a}}{kT}\right)
\end{equation}
and $A$ is assumed to be a constant. The activation energy is
determined from the measured data with a linear regression of
$\ln\sigma$ vs. $1/T$ for each gate voltage according to
$\ln\sigma=\ln A-E_{a}/kT$. The energetic difference between the
Fermi level $E_{F}$ of the sample and the valence band edge at the
insulator-semiconductor interface is approximated with the measured
activation energy $E_{a}(V_{g})$ of the field-effect conductivity
$\sigma$, i.e.
\begin{equation}
E_{a}\approx E_{V}-E_{F}-eV_{0}.
\end{equation}
$V_{0}=|V(x=0)|$ is the potential right at the
insulator-semiconductor interface. The $x$-direction is normal to
this interface. $E_{V}$ is the energy of the valence band edge far
from the insulator-semiconductor interface (at $x=d$, Fig.~3 in
Ref.~\onlinecite{KalbWL2008}). The underlying idea is the following:
a change of the gate voltage by $\Delta V_{g}$ leads to a shift of
the activation energy $E_{a}$ (i.e. of the effective Fermi level
$\tilde{E}_{F}=E_{F}+eV_{0}$ at the insulator-semiconductor
interface) by $\Delta \tilde{E}_{F}\approx\Delta E_{a}$. The change
in gate voltage $\Delta V_{g}$ corresponds to a total hole density
per unit area of $\Delta P=C_{i}\Delta V_{g}/e$. Then, the abrupt
approximation is made: the charge in the accumulation layer is
constant up to a distance $a$ from the insulator-semiconductor
interface and zero for larger distances. For the present method, it
is assumed that the parameter $a$ does not depend on gate
voltage.\cite{LangDV2004} Consequently, we have a change of the
volume hole density of $\Delta p=\Delta P/a$ close to the
insulator-semiconductor interface. By neglecting the free charge we
can estimate the trap density to be
\begin{equation} \label{equation-la1}
N=\Delta p/\Delta E_{a}=\frac{C_{i}}{ea}\left(\frac{\Delta
E_{a}}{\Delta V_{g}}\right)^{-1}.
\end{equation}
If one replaces the difference quotient in Eq.~\ref{equation-la1} by
the respective derivative we have the final result
\begin{equation} \label{equation-la2}
N(E)=\frac{C_{i}}{ea}\left(\frac{dE_{a}}{dV_{g}}\right)^{-1}.
\end{equation}
Consequently, the function $N(E)$ is calculated with
Eq.~\ref{equation-la2} and $N(E)$ is plotted as a function of the
energy $E=E_{a}(V_{g})\approx E_{V}-E_{F}-eV_{0}$.\cite{LangDV2004}

The band mobility $\mu_{0}$ may be estimated with
\begin{equation} \label{equation-la3}
\mu_{0}=\mu\exp\left(\frac{E_{a}}{kT}\right),
\end{equation}
by introducing the value of the measured activation energy $E_{a}$
and the field-effect mobility $\mu$ at a fixed and sufficiently high
gate voltage $V_{g}$.\cite{ButkoVY2003}

\begin{figure}
\includegraphics[width=0.90\linewidth]{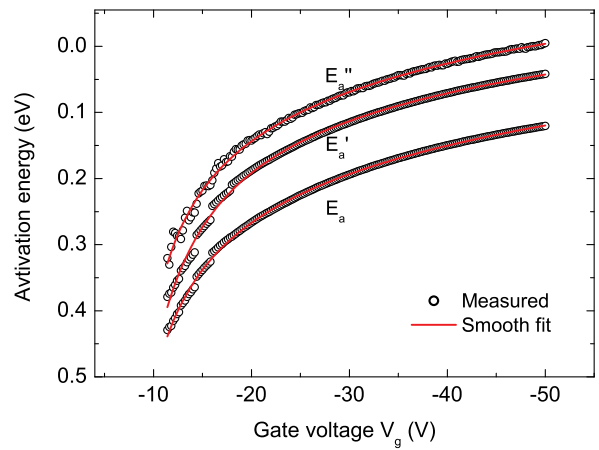}
\caption{\label{figure-activation} (Color online) Activation
energies $E_{a}$, $E_{a}'$ and $E_{a}''$ as determined with linear
regressions according to Eq.~\ref{equation-la0},
Eq.~\ref{equation-ika2} and Eq.~\ref{equation-fo4b}. The graph also
shows a smooth fit of the activation energies in each case (red
lines). There is a significant difference between $E_{a}$, $E_{a}'$
and $E_{a}''$.}
\end{figure}

We now give some specific details about the application of this
method to our data. The activation energy $E_{a}=E_{a}(V_{g})$ was
determined according to Eq.~\ref{equation-la0} and is shown in
Fig.~\ref{figure-activation}. The activation energy $E_{a}(V_{g})$
was then represented by a smooth fit (red line in
Fig.~\ref{figure-activation}) in order to suppress the noise in the
data. We used an effective accumulation layer thickness of
$a=7.5$\,nm.\cite{LangDV2004} The band mobility $\mu_{0}$ was
calculated with Eq.~\ref{equation-la3} for a high gate voltage of
$V_{g}=-50$\,V and T=299\,K.

\subsubsection{Method by Horowitz et al.}

Also for this method (Ref.~\onlinecite{HorowitzG1995}) the abrupt
approximation is made. However, in contrast to the method by
\textit{Lang et al.}, the present method allows for a gate-voltage
dependence of the effective accumulation layer thickness
$a=a(V_{g})$. As a consequence of the abrupt approximation, the
potential $V(x)$ in the organic semiconductor is given by
\begin{equation}
V(x)=V_{0}\left(1-\frac{x}{a}\right)^{2}
\end{equation}
with the interface potential
\begin{equation} \label{equation-ho0}
V_{0}\approx a\frac{C_{i}U_{g}}{2\epsilon_{0}\epsilon_{s}}.
\end{equation}
$U_{g}=|V_{g}-V_{FB}|$ is the gate voltage above the flatband
voltage $V_{FB}$. With the total hole density per unit area
$P=C_{i}U_{g}$ and Eq.~\ref{equation-ho0} we obtain an equation for
the total volume hole density $p$ which is
\begin{equation} \label{equation-ho1}
p=\frac{P}{a}\approx\frac{C_{i}^{2}U_{g}^{2}}{2\epsilon_{0}\epsilon_{s}eV_{0}}.
\end{equation}
Assuming again the abrupt approximation, it can be shown that:
\begin{equation} \label{equation-ho2}
eV_{0}=E_{V}-E_{F}-kT\ln\left(\frac{\mu_{0}N_{V}kT\epsilon_{0}\epsilon_{s}}{\mu
C_{i}^{2}U_{g}^{2}}\right).
\end{equation}
$\mu=\mu(V_{g})$ is the gate-voltage dependent field-effect mobility
as calculated with Eq.~\ref{equation-mu2}. $N_{V}$ is the effective
density of extended states. For each temperature, the trap DOS is
now calculated separately. To do so, a value of the product
$\mu_{0}N_{V}$ is assumed \textit{a priori } and the interface
potential $V_{0}$ is calculated with Eq.~\ref{equation-ho2}. We note
that this also requires an estimate of the difference between the
Fermi energy and the energy of the valence band edge far from the
insulator-semiconductor interface $E_{V}$, i.e. an estimate of
$E_{V}-E_{F}$. $V_{0}=V_{0}(V_{g})$ from Eq.~\ref{equation-ho2} is
used to calculate the volume hole density $p$ with
Eq.~\ref{equation-ho1}.
The Fermi function is approximated by a step function
(zero-temperature approximation). Its derivative then is a delta
function.
Consequently, the trap DOS is eventually obtained by numerically
differentiating the hole density $p$ from Eq.~\ref{equation-ho1}
with respect to the interface potential $V_{0}$ from
Eq.~\ref{equation-ho2}, i.e.
\begin{equation} \label{equation-ho3}
N(E)\approx\frac{1}{e}\frac{dp(V_{0})}{dV_{0}}.
\end{equation}
The trap densities from Eq.~\ref{equation-ho3} are finally plotted
as a function of the energy $E=E_{V}-E_{F}-eV_{0}$ as calculated
with Eq.~\ref{equation-ho2}. The trap DOS from the measurements at
different temperatures will generally not coincide at first. The
procedure is thus repeated for different values of the product
$\mu_{0}N_{V}$ until the trap DOS curves calculated from the data
taken at different temperatures, coincide.

The band mobility $\mu_{0}$ is calculated from the final parameter
$\mu_{0}N_{V}$ by assuming a value of the effective density of
extended states $N_{V}$. If $N_{V}$ is fixed, $\mu_{0}$ is the
adjustable parameter.

\begin{figure}
\includegraphics[width=0.90\linewidth]{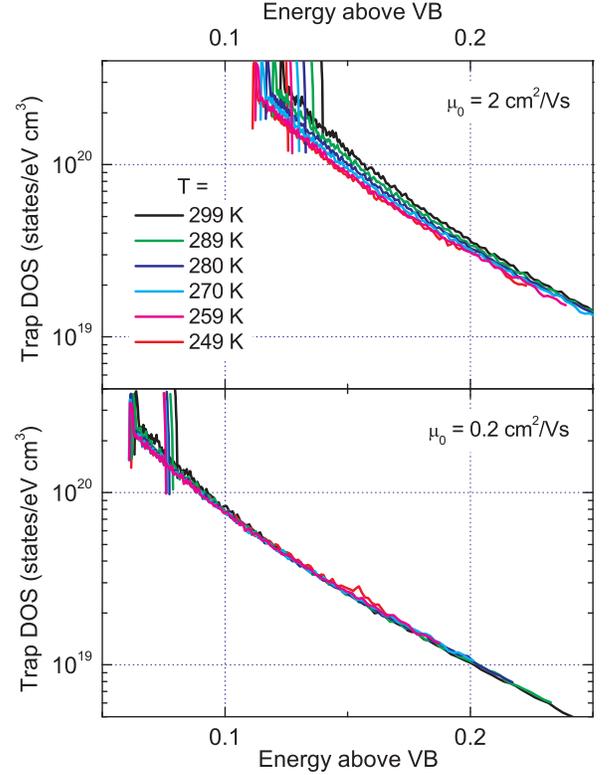}
\caption{\label{figure-horofit} (Color online) Trap densities
calculated with the method by \textit{Horowitz et al.} from the
measurements at different temperatures. The band mobility $\mu_{0}$
is an adjustable parameter in this method. Upper panel:
$\mu_{0}=2$\,cm$^{2}$/Vs, lower panel: $\mu_{0}=0.2$\,cm$^{2}$/Vs.
For $\mu_{0}=0.2$\,cm$^{2}$/Vs the trap densities from the
measurements at different temperatures coincide.}
\end{figure}

We proceed by giving specific details about the use of this method.
We estimated that $E_{V}-E_{F}=0.5$\,eV. In order to determine
$U_{g}=|V_{g}-V_{FB}|$, the flatband voltage $V_{FB}$ was taken to
be equal to the device onset voltage at room temperature. We thus
have $V_{FB}=-4.6$\,V. Moreover, the effective density of extended
states $N_{V}$ was assumed to be equal to the density of the
pentacene molecules, i.e. $N_{V}=3\times10^{21}$\,cm$^{-3}$. For
example, the trap densities were calculated from the measurements at
different temperatures with $\mu_{0}=2$\,cm$^{2}$/Vs and the result
is shown in the upper panel of Fig.~\ref{figure-horofit}. The trap
densities from the different temperatures do not coincide. The
procedure was repeated for different values of $\mu_{0}$. For
$\mu_{0}=0.2$\,cm$^{2}$/Vs the energetic distributions of traps do
coincide (lower panel of Fig.~\ref{figure-horofit}). This trap DOS
was taken as the final result along with the band mobility of
$\mu_{0}=0.2$\,cm$^{2}$/Vs.

\subsubsection{Method by Fortunato et al.}

This method (also called temperature method) is described in
Ref.~\onlinecite{FortunatoG1988} and \onlinecite{FogliettiV1999}.
The trap DOS is calculated with
\begin{equation} \label{equation-fo1}
N(E)=\frac{\epsilon_{0}\epsilon_{s}}{2e}\frac{\partial^{2}}{\partial
V_{0}^{2}}\left(\frac{dV(x)}{dx}\bigg|_{x=0}\right)^{2}.
\end{equation}
The electric field $dV/dx$ in Eq.~\ref{equation-fo1} may be written
as
\begin{equation} \label{equation-fo2}
\frac{dV(x)}{dx}\bigg|_{x=0}=\frac{\epsilon_{i}}{\epsilon_{s}}\frac{U_{g}-V_{0}}{l}.
\end{equation}
The interface potential $V_{0}$ in Eq.~\ref{equation-fo1} and
Eq.~\ref{equation-fo2} is obtained as described in the following.
First of all, the derivative of the field-effect conductivity can be
written as
\begin{equation} \label{equation-fo4X}
\frac{d\sigma}{dV_{g}}=\mu_{0}\frac{N_{V}\epsilon_{0}\epsilon_{i}}{lp(V_{0})}\exp\left(-\frac{E_{V}-E_{F}-eV_{0}}{kT}\right)
\end{equation}
and both the band mobility $\mu_{0}$ and the exponential factor
depend on temperature.\cite{FortunatoG1988} It can be shown, that
the total hole density $p(V_{0})$ varies much less with temperature
than the exponential term in
Eq.~\ref{equation-fo4X}.\cite{FortunatoG1988} As described in the
following, the calculation of a normalized field-effect conductivity
$\sigma'$ eliminates the temperature dependence due to the band
mobility $\mu_{0}$ in Eq.~\ref{equation-fo4X}.\cite{FogliettiV1999,
DeAngelisF2006} The field-effect mobility at RT and at each reduced
temperature $T$ are calculated with Eq.~\ref{equation-mu2}. Then,
the normalized field-effect conductivity $\sigma'$ is calculated for
each temperature according to
\begin{equation} \label{equation-fo3}
\sigma'=\sigma\frac{\mu^{RT}}{\mu^{T}}\approx
\sigma\frac{\mu_{0}^{RT}}{\mu_{0}^{T}}.
\end{equation}
$\mu^{RT}$ and $\mu^{T}$ in Eq.~\ref{equation-fo3} are the
field-effect mobilities at RT and at a reduced temperature $T$
evaluated at a fixed and sufficiently high gate voltage.
$\mu_{0}^{RT}$ and $\mu_{0}^{T}$ are the respective band
mobilities.\cite{FogliettiV1999, DeAngelisF2006} Clearly, the
derivative of the normalized field-effect conductivity may now be
written as
\begin{equation} \label{equation-fo4}
\frac{d\sigma'}{dV_{g}}=\mu_{0}^{RT}\frac{N_{V}\epsilon_{0}\epsilon_{i}}{lp(V_{0})}\exp\left(-\frac{E_{V}-E_{F}-eV_{0}}{kT}\right)
\end{equation}
and the exponential term is the only term with a temperature
dependence (the temperature dependence of $p(V_{0})$ is neglected).
Consequently, the activation energy $E_{a}''$ as determined with
linear regressions according to
\begin{equation} \label{equation-fo4b}
\frac{d\sigma'}{dV_{g}}\propto\exp\left(-\frac{E_{a}''}{kT}\right)
\end{equation}
is approximately equal to the difference between the Fermi energy
and the valence band edge at the insulator-semiconductor interface,
i.e.
\begin{equation} \label{equation-fo5}
E_{a}''\approx E_{V}-E_{F}-eV_{0}.
\end{equation}
Once $E_{a}''(V_{g})$ is known, the interface potential
$V_{0}(V_{g})$ can be calculated with Eq.~\ref{equation-fo5}
assuming \textit{a priori} a value for the difference $E_{V}-E_{F}$.
Then we can calculate the electric field with
Eq.~\ref{equation-fo2}. The result is finally introduced in
Eq.~\ref{equation-fo1} and the numerical differentiation with
respect to $V_{0}$ from Eq.~\ref{equation-fo5} eventually yields the
trap DOS as a function of energy $E=E_{a}''\approx
E_{V}-E_{F}-eV_{0}$.

It is instructive to consider the statistical shift in this context.
Both the Fermi energy $E_{F}$ and the interface potential $V_{0}$
depend on temperature. For the moment we assume that this
temperature-dependence is linear, i.e.
\begin{equation}\label{equation-stat1}
E_{V}-E_{F}=E_{V}-E_{F}^{0}+\alpha T
\end{equation}
and
\begin{equation} \label{equation-stat2}
eV_{0}=eV_{0}^{0}+(\alpha-\beta)T.
\end{equation}
$\alpha$ and $\beta$ are constants and $E_{F}^{0}$ and $V_{0}^{0}$
are the Fermi energy and the interface potential at $T=0$\,K. From
Eq.~\ref{equation-stat1} and Eq.~\ref{equation-stat2} we see that
$E_{V}-E_{F}-eV_{0}=E_{V}-E_{F}^{0}-eV_{0}^{0}+\beta T$. If this is
introduced into Eq.~\ref{equation-fo4} we have, within this linear
approximation, a constant prefactor $\exp(\beta/k)$ and the
activation energy $E_{a}''$ is in fact a better approximation of the
difference between the Fermi energy and the valence band edge at the
insulator-semiconductor interface at $T=0$\,K, i.e.
\begin{equation}
E_{a}''\approx E_{V}-E_{F}^{0}-eV_{0}^{0}.
\end{equation}
Nevertheless, the energy scale is not corrected for the statistical
shift for the present method. The trap DOS is simply plotted as a
function of $E_{a}''$. Neglecting the statistical shift is a common
feature of all analytical methods in the present comparison that
employ temperature-dependent measurements (all methods except for
the method by \textit{Gr\"unewald et al.}).

Again, we assumed that $E_{V}-E_{F}=0.5$\,eV and $V_{FB}=-4.6$\,V.
In order to determine the activation energy $E_{a}''$ with
Eq.~\ref{equation-fo4b}, the field-effect mobilities were calculated
with Eq.~\ref{equation-mu2} leading to a mobility that increases
monotonically with gate voltage at all temperatures. These functions
were evaluated at $V_{g}=-50$\,V. The mobilities at 299, 289, 280,
270, 259 and 249\,K respectively are $\mu=0.17$, 0.16, 0.15, 0.13,
0.11 and 0.09\,cm$^{2}$/Vs.\footnote{Alternatively,
Eq.~\ref{equation-revision3} could be used to calculate the
field-effect mobility. Eq.~\ref{equation-revision3} also leads to
mobilities that increase monotonically with gate voltage at all
temperatures. The absolute values of the mobilities are lower,
however. At $V_{g}=-50$\,V and with $V_{FB}=-4.6$\,V we have
mobilities of 0.082, 0.071, 0.062, 0.051, 0.041 and
0.032\,cm$^{2}$/Vs ($T=299$, 289, 280, 270, 259 and 249\,K).}
$E_{a}''$ is significantly different from the activation energy
$E_{a}$ (Fig.~\ref{figure-activation}). Again, the activation energy
$E_{a}''$ was represented by a smooth fit in order to suppress the
noise in the data (red line in Fig.~\ref{figure-activation}). The
smoothed function was used for the calculation of the trap DOS.

\subsubsection{Method by Gr\"unewald et al.}

This method does not require temperature-dependent
measurements.\cite{GrunewaldM1980, GrunewaldM1981, WeberK1982,
KalbWL20072} It allows to convert a single transfer characteristic
into the underlying density of states function which may be
advantageous in certain situations where temperature-dependent
measurements are not possible.\cite{KalbWL20072} In addition, it is
not necessary to consider the temperature dependence of the Fermi
energy, the interface potential or the band mobility. Moreover, the
method is not based on the abrupt approximation. The interface
potential $V_{0}$ as a function of gate voltage is calculated from
\begin{eqnarray} \label{equation-gr0}
\exp \bigg(\frac{eV_{0}}{kT}\bigg)-\frac{eV_{0}}{kT}-1 \nonumber
\\
=\frac{e}{kT}\frac{\epsilon_{i}d}{\epsilon_{s}l\sigma_{0}}\bigg[U_{g}\sigma(U_{g})-\int_{0}^{U_{g}}\sigma(\tilde{U_{g}})d\tilde{U_{g}}\bigg].
\end{eqnarray}
For each gate voltage, Eq.~\ref{equation-gr0} is numerically
evaluated using the measured field-effect conductivity $\sigma$
(Eq.~\ref{equation-sig2}). Eventually, we have the complete function
$V_{0}=V_{0}(V_{g})$. The total hole density $p$ can be calculated
with $V_{0}$ according to
\begin{equation} \label{equation-gr2}
p(V_{0})=\frac{\epsilon_{0}\epsilon_{i}^{2}}{\epsilon_{s}l^{2}e}U_{g}\left(\frac{dV_{0}}{dU_{g}}\right)^{-1}.
\end{equation}
Within the zero-temperature approximation the trap DOS $N(E)$ can
then be written as
\begin{equation} \label{equation-gr1}
N(E)\approx\frac{1}{e}\frac{dp(V_{0})}{dV_{0}}.
\end{equation}
This means that we do a numerical differentiation of the hole
density from Eq.~\ref{equation-gr2} with respect to the interface
potential $V_{0}$ from Eq.~\ref{equation-gr0}. The trap DOS can be
plotted as a function of the energy $E=eV_{0}$, i.e. as a function
of the energy relative to the Fermi energy $E_{F}$. It can also be
plotted as a function of the energy $E=E_{V}-E_{F}-eV_{0}$ which
requires the difference $E_{V}-E_{F}$ to be estimated.

For the present method, the gated four-terminal measurement at
$T=299$\,K was considered. Again, we assumed $E_{V}-E_{F}=0.5$\,eV
and $V_{FB}=-4.6$\,V.

\subsubsection{Method I by Kalb et al.}

The free hole density $P_{free}$ per unit area is written as
\begin{equation} \label{equation-ka0}
P_{free}\approx
a(V_{g})N_{V}\exp\left(-\frac{E_{V}-E_{F}-eV_{0}}{kT}\right)
\end{equation}
with
\begin{equation} \label{equation-ka00}
a(V_{g})=\frac{m}{2m-1}\frac{2kT\epsilon_{0}\epsilon_{s}}{eC_{i}U_{g}}.
\end{equation}
$a(V_{g})$ is the effective thickness of the accumulation layer.
$E_{0}=kT_{0}$ is the slope of the trap DOS and
$m=T_{0}/T$.\cite{KalbWL2008} Since the field-effect conductivity
can be written as
\begin{equation} \label{equation-ka000}
\sigma=e\mu_{0}P_{free},
\end{equation}
the difference $E_{V}-E_{F}-eV_{0}$ is approximated by the
activation energy $E_{a}(V_{g})$ of the field-effect conductivity
$\sigma$. $E_{a}$ is determined with linear regressions from the
measured data according to Eq.~\ref{equation-la0}. This procedure
implies, that the temperature dependence of the mobility $\mu_{0}$
as well as the temperature dependence of the effective accumulation
layer thickness $a$ are negligible compared to the exponential
temperature dependence. By substituting $dV_{0}=-dE_{a}/e$ in
Eq.~\ref{equation-gr1} and Eq.~\ref{equation-gr2}, we finally have
the trap DOS
\begin{equation} \label{equation-ka2}
N(E)\approx\frac{d}{dE_{a}}\left[\frac{\epsilon_{0}\epsilon_{i}^{2}}{\epsilon_{s}l^{2}}U_{g}\left(\frac{dE_{a}}{dU_{g}}\right)^{-1}\right]
\end{equation}
as a function of the energy $E=E_{a}(V_{g})\approx
E_{V}-E_{F}-eV_{0}$.

The band mobility $\mu_{0}$ can be estimated with
\begin{equation} \label{equation-ka3}
\mu_{0}=\sigma/(eP_{free}),
\end{equation}
where $\sigma$ is the measured field-effect conductivity and
$P_{free}$ is calculated according to
Eq.~\ref{equation-ka0}.\cite{KalbWL2008}

We give some specific details about the use of this method: We used
$V_{FB}=-4.6$\,V. The trap DOS was calculated with
Eq.~\ref{equation-ka2} from the smooth fit of the activation energy
$E_{a}$ in Fig.~\ref{figure-activation}. The parameter $m=T_{0}/T$
in Eq.~\ref{equation-ka00} is only relevant for the calculation of
the band mobility $\mu_{0}$ with Eq.~\ref{equation-ka3}. To obtain
this parameter, an exponential function $N(E)=N_{0}\exp(-E/E_{0})$
was fitted to the trap DOS that had previously been obtained with
Eq.~\ref{equation-ka2}. This gave $E_{0}=kT_{0}=60$\,meV and thus
$m=2.33$ at $T=299$\,K. The band mobility was calculated for a gate
voltage of $V_{g}=-50$\,V and $T=299$\,K.

\subsubsection{Method II by Kalb et al.}

As suggested by \textit{Fortunato et al.}, the temperature
dependence of the band mobility $\mu_{0}$ can be eliminated by
calculating a normalized field-effect conductivity $\sigma'$ for
each temperature according to Eq.~\ref{equation-fo3}.

In order to improve upon the method by \textit{Kalb et al.}, the
normalized activation energy $E_{a}'(V_{g})$ is determined for each
gate voltage with a linear regression according to
\begin{equation} \label{equation-ika2}
\sigma'(V_{g})\propto\exp\left(-\frac{E_{a}'}{kT}\right).
\end{equation}
$\sigma'$ in Eq.~\ref{equation-ika2} is the normalized field-effect
conductivity according to Eq.~\ref{equation-fo3}. $E_{a}'$ is a
better approximation of the difference between the Fermi energy and
the valence band edge, i.e. we now have
\begin{equation}
E_{a}'\approx E_{V}-E_{F}-eV_{0}.
\end{equation}
$E_{a}'$ is now used instead of $E_{a}$ in Eq.~\ref{equation-ka2},
i.e. the trap DOS is finally calculated with
\begin{equation}
N(E)\approx\frac{d}{dE_{a}'}\left[\frac{\epsilon_{0}\epsilon_{i}^{2}}{\epsilon_{s}l^{2}}U_{g}\left(\frac{dE_{a}'}{dU_{g}}\right)^{-1}\right].
\end{equation}
It is plotted as a function of the energy $E=E_{a}'\approx
E_{V}-E_{F}-eV_{0}$.

\subsubsection{Influence of the choice of parameters}

\begin{figure}
\includegraphics[width=0.90\linewidth]{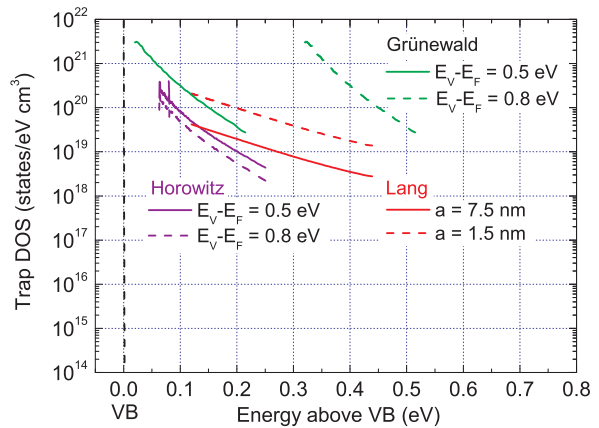}
\caption{\label{figure-errors} (Color online) The choice of certain
parameters in the analytical methods can have a significant effect
on the trap DOS. For the method by \textit{Lang et al.}, an
effective accumulation layer thickness of $a=1.5$\,nm (dashed red
line) instead of 7.5\,nm (full red line) leaves the slope of the
trap DOS unchanged but leads to a significant increase of the
overall trap densities. For the method by \textit{Horowitz et al.},
the use of $E_{V}-E_{F}=0.8$\,eV (dashed violet line) instead of
$E_{V}-E_{F}=0.5$\,eV (full violet line) results in a rather small
change in the magnitude of the trap densities. The method by
\textit{Gr\"unewald et al.} very sensitively depends on the choice
of $E_{V}-E_{F}$: using $E_{V}-E_{F}=0.8$\,eV (dashed green line)
instead of $E_{V}-E_{F}=0.5$\,eV (full green line) results in a
parallel shift of the trap DOS by 0.3\,eV.}
\end{figure}

We also investigated how the choice the effective accumulation layer
thickness $a$ and the difference between the Fermi level and the
valence band edge far from the insulator-semiconductor interface
$E_{V}-E_{F}$ effect the final result. The effective accumulation
layer thickness $a$ needs to be fixed for the method by \textit{Lang
et al.} Clearly, this choice significantly affects the trap DOS
(Fig.~\ref{figure-errors}). For the method by \textit{Horowitz et
al.}, $E_{V}-E_{F}=0.5$\,eV was chosen. We repeated the calculations
for $E_{V}-E_{F}=0.8$\,eV and again, the trap densities from the
measurements at different temperatures were found to coincide with a
parameter of $\mu_{0}=0.2$\,cm$^{2}$/Vs. The results are compared in
Fig.~\ref{figure-errors}. The slope of the trap DOS calculated for
$E_{V}-E_{F}=0.5$\,eV and 0.8\,eV are almost identical in both
cases, but the trap densities are reduced to some extend due to the
larger value of $eV_{0}$ in the denominator of
Eq.~\ref{equation-ho1}. Also for the method by \textit{Fortunato et
al.}, the parameter $E_{V}-E_{F}$ needs to be known. However, this
method is not sensitive to the choice of this parameter:
calculations for $E_{V}-E_{F}=0.8$\,eV give essentially the same
result. Clearly, for the method by \textit{G\"unewald et al.}, the
guess of $E_{V}-E_{F}$ has a significant influence on the trap DOS
since the energy scale is given by $E_{V}-E_{F}-eV_{0}$ and only
$eV_{0}$ is known. Therefore, the choice of $E_{V}-E_{F}=0.8$\,eV
instead of 0.5\,eV  leads to a parallel shift of the trap DOS by
0.3\,eV along the energy scale as shown in Fig.~\ref{figure-errors}.
For the method by \textit{Kalb et al.} we neither need to make an
assumptions about $E_{V}-E_{F}$ nor about the effective accumulation
layer thickness $a$.

The necessity to choose the effective density of extended states
($N_{V}=3\times10^{21}$\,cm$^{-3}$ in the present study) leads to an
uncertainty in the absolute vale of the band mobility. The volume
density of extended states is given by the density of the pentacene
molecules ($\approx3\times10^{21}$\,cm$^{-3}$) multiplied by two due
to the spin degree of freedom.\cite{ScheinertS2004, PernstichKP2006,
PernstichKP2008} However, a value of
$\approx6\times10^{21}$\,cm$^{-3}$ is likely to overestimate the
effective density of extended states $N_{V}$. The underlying
spectral density of extended states (in cm$^{-3}$eV$^{-1}$) should
drop close to the valence band edge. The extended states close to
the valence band edge are the most important contribution to $N_{V}$
though. Therefore, the volume density of molecules without the
degeneracy factor (Ref.~\onlinecite{KalbWL2008}) or even half the
molecular density (Ref.~\onlinecite{HorowitzG1995}) have been used
as approximations of $N_{V}$. In the case of pentacene, a value as
low as $N_{V}=1\times10^{21}$\,cm$^{-3}$ has also been
used.\cite{ScheinertS2007} From the method by \textit{Horowitz et
al.} it appears, that $N_{V}$ cannot be higher than
$3\times10^{21}$\,cm$^{-3}$. A band mobility of
$\mu_{0}=0.2$\,cm$^{2}$/Vs was calculated from the product
$\mu_{0}N_{V}$ with $N_{V}=3\times10^{21}$\,cm$^{-3}$: a larger
$N_{V}$ would lead to a band mobility lower than the field-effect
mobility which would not be a reasonable result.

\subsection{\label{subsection-computer} Computer simulation of the transfer characteristics}

For the modeling of the transfer characteristics, we used the
Matlab\circledR-based program developed by \textit{Oberhoff et
al}.\cite{OberhoffD2007} The essence of this approach is that the
program calculates the transfer characteristics from the trap DOS,
the spectral density of extended states and the band mobility
$\mu_{0}$. It is assumed that the valence band has a rectangular
shape, i.e. the density of extended states (in cm$^{-3}$eV$^{-1}$)
is constant anywhere in the valence band.\cite{OberhoffD2007} The
program can simulate the transfer characteristic at any temperature
$T$ as long as the band mobility $\mu_{0}$ at this temperature is
also fixed \textit{a priori}. The full Fermi-Dirac statistics is
included.\cite{OberhoffD2007} This means that, in contrast to the
analytical methods, the Fermi function is not approximated and the
temperature-dependence of the Fermi energy $E_{F}$ is not neglected.
In Fig.~\ref{figure-tempTFT} we show simulated transfer
characteristics (red lines) closely matching the measured data
(symbols). The trap DOS from which these transfer characteristics
were calculated is also shown in Fig.~\ref{figure-compmethods}
(light blue line). Table~\ref{table-compmethods} lists the
respective parameters including the band mobility at RT.

We now give some more specific details about the simulations. The
program allows for a consideration of parasitic resistances at the
source and drain contacts. For the present simulations, we have
however assumed negligible contact resistances. This is supported by
the gated four-terminal measurements, which show that the total
contact resistance is significantly lower than the channel
resistance at all temperatures. More specifically at $V_{g}=-50$\,V
and $T=299$\,K, the channel resistance is about ten times larger
than the contact resistance and still five times larger than the
contact resistance at $V_{g}=-50$\,V and $T=249$\,K. We use
10$^{22}$\,cm$^{-3}$eV$^{-1}$ for the density of extended states in
the rectangular band. In essence, this is an approximation of the
density of extended states close to the valence band edge since only
these states are of importance for the charge transport. The value
is lower by a factor of two compared to the volume density of
extended states ($6\times10^{21}$\,cm$^{-3}$) divided by the
bandwidth (0.3\,eV, Ref.~\onlinecite{NabokD2007}). The reduced value
accounts for a drop of the spectral density of extended states close
to the valence band edge in analogy to the choice of
$N_{V}=3\times10^{21}$\,cm$^{-3}$ for the analytical methods. To
obtain a good fit of the transfer characteristics at all
temperatures, it was necessary to allow for a temperature dependence
of the band mobility $\mu_{0}$. For the fit in
Fig.~\ref{figure-tempTFT}, the band mobilities $\mu_{0}$ at 299,
289, 280, 270, 259 and 249\,K were fixed at 0.32, 0.28, 0.25, 0.21,
0.17 and 0.14\,cm$^{2}$/Vs, respectively.

\section{\label{section-discussion} Discussion}

We begin with a discussion of the results in
Fig.~\ref{figure-compmethods} and Table~\ref{table-compmethods}. The
estimate from the subthreshold swing assumes the trap densities not
to depend on energy. It is derived from the subthreshold swing and
can thus be regarded as a rough estimate of the density of traps
slightly above the Fermi energy $E_{F}$. It is however gratifying to
note the agreement between the trap DOS derived from the
subthreshold swing (valid for $E\approx E_{F}$) and the lowest
values of the trap DOS from the other methods. All other methods
result in a trap DOS that increases somewhat faster than
exponentially with energy. From Fig.~\ref{figure-compmethods} and
Table~\ref{table-compmethods} we see, that the choice of the method
to calculate the trap DOS has a considerable effect on the final
result.

The method by \textit{Kalb et al.} gives a good estimate of the
slope of the trap DOS, but leads to an overestimation of the overall
magnitude of the trap densities, i.e. the parameter $N_{0}$. This is
because the temperature dependence of the band mobility $\mu_{0}$ in
Eq.~\ref{equation-ka000} is neglected compared to the exponential
temperature dependence (Eq.~\ref{equation-ka0}). The improved method
by \textit{Kalb et al.} (method II) uses the activation energy
$E_{a}'$ instead of $E_{a}$, and $E_{a}'$ is calculated from the
normalized field-effect conductivity at each temperature
(Eq.~\ref{equation-ika2}). This means that the temperature
dependence of the band mobility is properly taken account of. The
correction has a considerable effect on the overall magnitude of the
trap densities. The improved method leads to a trap DOS that is in
much better agreement with the other methods and in particular with
the result from the simulations. This means that the temperature
dependence of the band mobility $\mu_{0}$ should generally not be
neglected when calculating the trap DOS with an analytical method.
Method II by \textit{Kalb et al.} is similar to the method by
\textit{Fortunato et al}. It is easier to be used but is based on
additional simplifications. The difference between the method by
\textit{Fortunato et al.} and the method II by \textit{Kalb et al.}
is the use of the activation energy $E_{a}''$ of the derivative
$d\sigma'/dV_{g}$ instead of the activation energy $E_{a}'$ of
$\sigma'$. Moreover, $V_{0}$ in Eq.~\ref{equation-fo2} is not
neglected by \textit{Fortunato et al.}, contrary to the method by
\textit{Kalb et al.} Neglecting $V_{0}$ does not lead to significant
differences for sufficiently high gate voltages since the interface
potential is typically less than 0.5\,V. On the other hand, from
Fig.~\ref{figure-activation} we see that there are significant
differences between $E_{a}'$ and $E_{a}''$. The difference between
the two methods, therefore, is almost exclusively due to the use of
$E_{a}''$ instead of $E_{a}'$.
The method II by \textit{Kalb et al.} corrects for the
temperature-dependence of the band mobility but neglects the
temperature-dependence of $a(V_{g})$ (Eq.~\ref{equation-ka00})
against the exponential temperature-dependence in
Eq.~\ref{equation-ka0}. This still is a source of error. For the
present example we have $mkT/(2m-1)=16.41$\,meV at $T=299$,K and
$mkT/(2m-1)=13.07$\,meV at $T=249$\,K, i.e. a ratio of 1.26. This
should be compared to $\mu^{RT}/\mu^{T}=1.9$
($\mu^{RT}=0.17$\,cm$^{2}$/Vs and $\mu^{T}=0.09$\,cm$^{2}$/Vs at
$T=249$\,K).
The method by \textit{Fortunato et al.} is in excellent agreement
with the result from the computer simulations.

The method by \textit{Lang et al.} leads to a significant error in
the slope of the trap DOS. This is mainly due to the assumption of a
gate-voltage independent effective accumulation layer thickness $a$.
If we allow for a gate-voltage dependent effective accumulation
layer thickness in the context of the abrupt approximation, the
effective thickness $a(V_{g})$ decreases with increasing gate
voltage (Eq.~\ref{equation-ka00}). As a consequence, the assumption
of a constant thickness $a$ in the denominator of
Eq.~\ref{equation-la2} leads to an overestimation of the trap
density at low gate voltages (at energies far from the valence band
edge) and to an underestimation of the trap density at high gate
voltages (at energies close to the valence band edge).

From Fig.~\ref{figure-errors} we see that some analytical methods do
not lead to an unambiguous result. In principle, the difference
$E_{V}-E_{F}$ may be approximated with the activation energy near
the flatband condition. However, this activation energy can often
not be measured because the off-current of an organic field-effect
transistor is often due to experimental limitations and is not
related to the conductivity of the organic semiconductor. If
$E_{V}-E_{F}$ is not experimentally accessible, we have an
uncertainty in the trap densities using the methods by
\textit{Horowitz et al.} and \textit{Gr\"unewald et al.} as shown in
Fig.~\ref{figure-errors}. Moreover, Fig.~\ref{figure-errors} shows
the significant dependence of the result from the method by
\textit{Lang et al.} on the choice of the constant effective
accumulation layer thickness $a$. The methods by \textit{Fortunato
et al.} and by \textit{Kalb et al.} do not lead to these
uncertainties.

The analytical methods approximate the Fermi function to a step
function for the trapped holes and use Boltzmann's approximation for
the free holes. The temperature dependencies of the Fermi energy
$E_{F}$ and the interface potential $V_{0}$ are also neglected.
These assumptions appear to be less restrictive because the trap
distributions from most analytical methods are in good agreement
with the result from the simulations which do not involve these
assumptions.

Most analytical methods lead to a band mobility $\mu_{0}$ that is
comparable to the value of $\mu_{0}$ from the simulations. It is
important to note that the band mobility $\mu_{0}$ from most methods
is only slightly higher than the field-effect mobility $\mu$ at high
gate voltages, i.e. $\mu\approx0.2$\,cm$^{2}$/Vs at $V_{g}=-50$\,V
and $T=299$\,K for this sample. Since $\mu=(P_{free}/P)\mu_{0}$ this
means that even in samples with an increased trap density (TFT's),
most of the gate-induced holes are free.

For the method by \textit{Fortunato et al.} and for method II by
\textit{Kalb et al.}, the activation energy ($E_{a}''$ or $E_{a}'$)
is calculated from the normalized field-effect conductivity at each
temperature. The field-effect conductivity is normalized to the
field-effect mobilities at high gate voltages $\mu^{RT}/\mu^{T}$.
This ratio is an approximation of the ratio of the respective band
mobilities $\mu_{0}^{RT}/\mu_{0}^{T}$. For example, the field effect
mobility at $V_{g}=-50$\,V is $\mu^{RT}=0.17$\,cm$^{2}$/Vs at
$299$\,K and $\mu^{T}=0.09$\,cm$^{2}$/Vs at $249$\,K. This gives a
ratio of $\mu^{RT}/\mu^{T}=1.9$.\footnote{If the mobilities are
calculated with Eq.~\ref{equation-revision3}, we get a ratio of
$\mu^{RT}/\mu^{T}=2.6$.} From the simulations we have a band
mobility of $\mu_{0}^{RT}=0.32$\,cm$^{2}$/Vs at $299$\,K and
$\mu_{0}^{T}=0.14$\,cm$^{2}$/Vs at $249$\,K. The ratio of the band
mobilities thus is $\mu_{0}^{RT}/\mu_{0}^{T}=2.3$. These two ratios
are very similar, indeed. This further supports the correction of
the field-effect conductivity as suggested by \textit{Fortunato et
al}.

We also note that from the simulations we have a band mobility
$\mu_{0}$ that decreases as the sample is cooled down. This may
indicate that the trap-free transport process is a hopping transport
but may also be limited by a thermally activated process at the
grain boundaries.

Finally, we recall that the trap DOS was calculated from transistors
with a rather small gate capacitance and we thus have relatively
large operating voltages. If transistors with a high gate
capacitance are to be used in order to quantify the trap DOS great
caution is required. Transistors with a sufficiently high gate
capacitance can be operated at gate voltages of only a few volts
(e.g. $2-3$\,V).\cite{KlaukH2007} This is comparable to the
magnitude of the interface potential $V_{0}$ ($\leq0.5$\,V).
However, all methods apart from the method by \textit{Fortunato et
al.} assume that the total charge per unit area can be approximated
according to $C_{i}(V_{g}-V_{FB}-V_{0})\approx C_{i}(V_{g}-V_{FB})$.
Significant errors are to be expected if $V_{0}$ is neglected for
low-voltage operating transistors. The method by \textit{Fortunato
et al.} does not neglect $V_{0}$ and could thus be used in this
scenario. For a low-voltage operating transistor, the choice of the
parameter $E_{V}-E_{F}$ for the method by \textit{Fortunato et al.}
is however expected to be a source of ambiguity because $V_{0}$ can
no longer be neglected in the numerator of Eq.~\ref{equation-fo2}.

\vspace{0.5cm}

\section{Summary and conclusions}

Several different methods were used to quantify the spectral density
of localized states in a pentacene-based organic thin-film
transistor. The trap DOS derived from the simple formula for the
subthreshold swing is in rather good agreement with the lowest
values of the trap DOS from the other, more sophisticated methods.
Most methods result in an almost exponential trap DOS close to the
valence band edge with a typical slope of $50$\,meV. We find that
the choice of the method to calculate the trap DOS has a
considerable effect on the final result. More specifically, two
assumptions lead to significant errors in the trap DOS. First,
neglecting the temperature dependence of the band mobility can lead
to a rather large overestimation of the trap densities. Secondly,
the assumption of a gate-voltage independent effective accumulation
layer thickness results in a significant underestimation of the
slope of the trap DOS. A general conclusion of this study is that it
is necessary to consider the specific deviations of a given
calculation method if one compares energetic distributions of trap
states from organic field-effect transistors evaluated by different
groups with different methods.

The methods by \textit{Fortunato et al.} and the method II by
\textit{Kalb et al.} do not lead to ambiguities due to the choice of
parameters, and this constitute a significant advantage. The
computer simulations do not approximate the Fermi function and may
therefore be seen as the most reliable result. Simulating the
transfer characteristics at various temperatures can, however, be a
time consuming endeavor due to the large number of possibilities to
fix the trap DOS and the band mobilities. While all methods have
their advantages and disadvantages, the method by \textit{Fortunato
et al.} is relatively easy to use and gives an unambiguous result in
excellent agreement with the computer simulations.

\begin{acknowledgments}
The authors thank Simon Haas and Andreas Reinhard for the growth of
pentacene crystals and Thomas Mathis and Kurt Pernstich for valuable
discussions.
\end{acknowledgments}



\end{document}